\documentclass{elsart}

\usepackage{natbib}
\usepackage{graphicx}

\usepackage{amssymb}

\newcommand{\eqb}{\begin{eqnarray}}
\newcommand{\eqe}{\end{eqnarray}}
\newcommand{\diff}{{\rm d}}
\newcommand{\mnras}{MN}
\newcommand{\apj}{ApJ}
\newcommand{\apjl}{ApJ}
\newcommand{\aap}{A\&A}
\newcommand{\apss}{Astrophys.\ \& Space Sci.}

\begin{document}

\begin{frontmatter}



\title{Dissipation in Pulsar Winds}


\author{J. G. Kirk}

\address{Max-Planck-Institut f\"ur Kernphysik,
Postfach 10 39 80, 69029 Heidelberg, Germany}

\begin{abstract}
I review the constraints placed on relativistic pulsar winds
by comparing optical and X-ray images of the inner
Crab Nebula on the one hand with two-dimensional MHD simulations 
on the other. The various proposals in the literature 
for achieving the low magnetisation required at the inner edge of
the Nebula, are then discussed, emphasising that of 
dissipation in the striped-wind picture. The possibility of 
direct observation of the wind is examined. 
Based on the predicted orientation of the polarisation 
vector, I outline a new argument suggesting that the off-pulse 
component of the optical emission of the Crab pulsar originates 
in the wind.
\end{abstract}


\end{frontmatter}

\section{Introduction}
\label{intro}



Observations of pulsar nebulae have improved dramatically over the past few
years. In particular, the X-ray \citep{weisskopfetal00} and optical
\citep{hesteretal95,hesteretal02} images of the Crab Nebula have revealed
intriguing insights into the dynamics of this object 
that have motivated detailed theoretical
studies. These, in turn, have sharpened the constraints on the possible physical
conditions at the inner boundary of the Nebula, which encloses both 
the pulsar and
an apparently empty region surrounding it. This paper concentrates on the
physics of the apparently empty region, that I will call 
the pulsar {\em wind},
because it contains out-flowing material, moving, 
presumably, at supersonic velocity. However, it
also contains large amplitude waves of the same period as the pulsar
\citep{reesgunn74, kundtkrotscheck80} and 
a substantial dc component of magnetic field \citep{pacini68}.

Whereas the outer boundary of a pulsar wind is, at least in some cases,
directly accessible to observation, 
constraints on the physical conditions where the wind is launched are 
difficult to find. A common scenario is that the pulsar magnetosphere  
drives an almost radial wind that becomes supersonic at some point not too far
outside the light cylinder and that
contains a current sheet separating the two
magnetic hemispheres defined by the dipole component of the field at the
stellar surface. However, although this is an attractive model to work on,
alternative pictures are possible \citep[e.g.,][]{kuijpers01}, because 
our knowledge of the magnetosphere is incomplete.  
Even the term {\em magnetosphere} itself can cause confusion. It 
is generally used to identify a region in which
external electromagnetic fields dominate the dynamics of a plasma. 
For pulsars, this
certainly applies inside the light cylinder. 
But the
definition becomes ambiguous further out, where the magnetic field weakens and
the outflow accelerates. This is because electromagnetic fields 
may still 
dominate the energy flux even in a supermagnetosonic flow, in which the 
inertia of the plasma is important. 
Here, I use the term magnetosphere to 
refer to regions close to the neutron star where
the bulk radial flow is still submagnetosonic, including, of course, those
parts within the pulsar light-cylinder that corotate with the star.
Understood in this sense, it is generally thought that 
the wind is dark, but the magnetosphere 
is responsible for the pulsed emission detected, in the case of the Crab,
from the radio to gamma-ray bands as a point source apparently located 
at the position of the neutron star. As discussed in
Section~\ref{observability} this is not necessarily the case.

In this paper I will briefly review the constraints that can be placed 
on the physics of the pulsar wind 
by MHD simulations of the Crab Nebula. 
I will then consider current theoretical ideas on how these
constraints could be met, and finally look at the prospects for 
observing the wind directly.

\section{MHD simulations}
Axisymmetric, relativistic 
MHD simulations of the wind and the bubble it 
inflates downstream of the termination shock have been performed by  
\citet{komissarovlyubarsky03,komissarovlyubarsky04} and
\citet{delzannaetal04}. 
The results are broadly similar. Provided the energy flux in the wind is
assumed to be concentrated towards the equatorial plane, the termination shock
is highly oblate --- an aspect that had been deduced using analytical
arguments by \citet{lyubarsky02} and 
\citet{bogovalovkhangoulian02b,bogovalovkhangoulian02a}. 
If, in addition, 
the magnetisation parameter $\sigma=B^2/(4\pi w)$ (here 
$B$ is the magnetic
field in the plasma rest frame and $w$ the proper enthalpy density)
is tuned to the correct
value, then 
the magnetic pinch in the shocked plasma collimates the flow, as predicted
\citep{lyubarsky02}, and
a torus-plus-jet like structure similar in appearance
to that seen in the 
optical and X-ray images of the Crab Nebula is reproduced.
The anisotropy of the energy flux and the tuning of the magnetisation are
critical; the simulations are less sensitive to other 
parameters such as the Lorentz factor of the flow and the angular 
distribution of the particle flux.

The anisotropy of the energy flux 
should follow from the boundary conditions on the
magnetic field at the stellar surface. But our knowledge of this connection is
scanty. The only known exact solution is that of the force-free 
split monopole
\citep{michel73}. In this case, spherical symmetry of the magnetic field 
at small radius yields an
energy flow per solid angle, $\diff L/\diff\Omega$, that is concentrated into the equatorial plane, 
with a dependence on colatitude $\theta$ given by
\eqb
{\diff L\over\diff\Omega}&\propto& \sin^2\theta
\label{monopoledistribution}
\eqe
The poloidal field in this solution is purely radial, and there are no 
regions of closed field. Nevertheless, a non-axisymmetric version can be
constructed by simply inclining the plane used to \lq\lq split\rq\rq\ the
monopole \citep{bogovalov99}.
In a realistic case, it is expected that the closed field line regions fill
much of the magnetosphere inside the light cylinder [see the discussion in
\citet{contopouloskazanasfendt99}, 
\citet{uzdensky03} and \citet{goodwinetal04}]. Consequently, there is no 
reason to suppose that the angular distribution given by 
Eq.~(\ref{monopoledistribution}) is a good approximation. 
Nevertheless, all simulations
published so far start off with essentially this distribution.

There is less consensus concerning the magnetisation parameter $\sigma$. This
quantity is, in principle, also a function of angle, 
since it describes the way in which
the energy flux at a particular colatitude 
is divided between Poynting flux and particle-born energy.
In the exact force-free solution $\sigma$ is formally infinite. 
But a simulation
with large $\sigma$ would fail to look anything like the 
Crab Nebula. In fact, for $\sigma$ greater than about 1\% in the equatorial
region, the jet appears too fast and strong and the
equatorial outflow is suppressed. For smaller $\sigma$, 
some of the detailed properties of the outflow 
match observed features such as the central
knot \citep{hesteretal95} quite convincingly \citep{komissarovlyubarsky04}. 
Thus, it appears that the
magnetisation parameter must be very small in the equatorial wind. This is 
a well-known and puzzling property, as discussed below. It is 
sometimes termed the \lq\lq $\sigma$ paradox\rq\rq, 
and was already recognised in spherically symmetric models of the Crab Nebula 
\citep{kennelcoroniti84a}. On the other hand, if $\sigma$ is everywhere small, 
the two-dimensional simulations
resemble the purely hydrodynamical case, resulting in 
a \lq\lq lonely torus\rq\rq\ \citep{komissarovlyubarsky04}.

\begin{figure}
\begin{center}
\caption{\label{prlfigure}
The striped pulsar wind. A magnetic dipole embedded in
the star at an oblique angle to the rotation axis introduces field
lines of both polarities into the equatorial plane. The figure shows the 
intersection of the current sheet 
separating these regions and the equatorial plane. 
In the inset, 
an almost planar portion of this sheet (dashed line) is shown,
together with the magnetic  
field lines, assuming they undergo reconnection.}
\includegraphics[bb=0 0 759 560,width=8.5 cm]{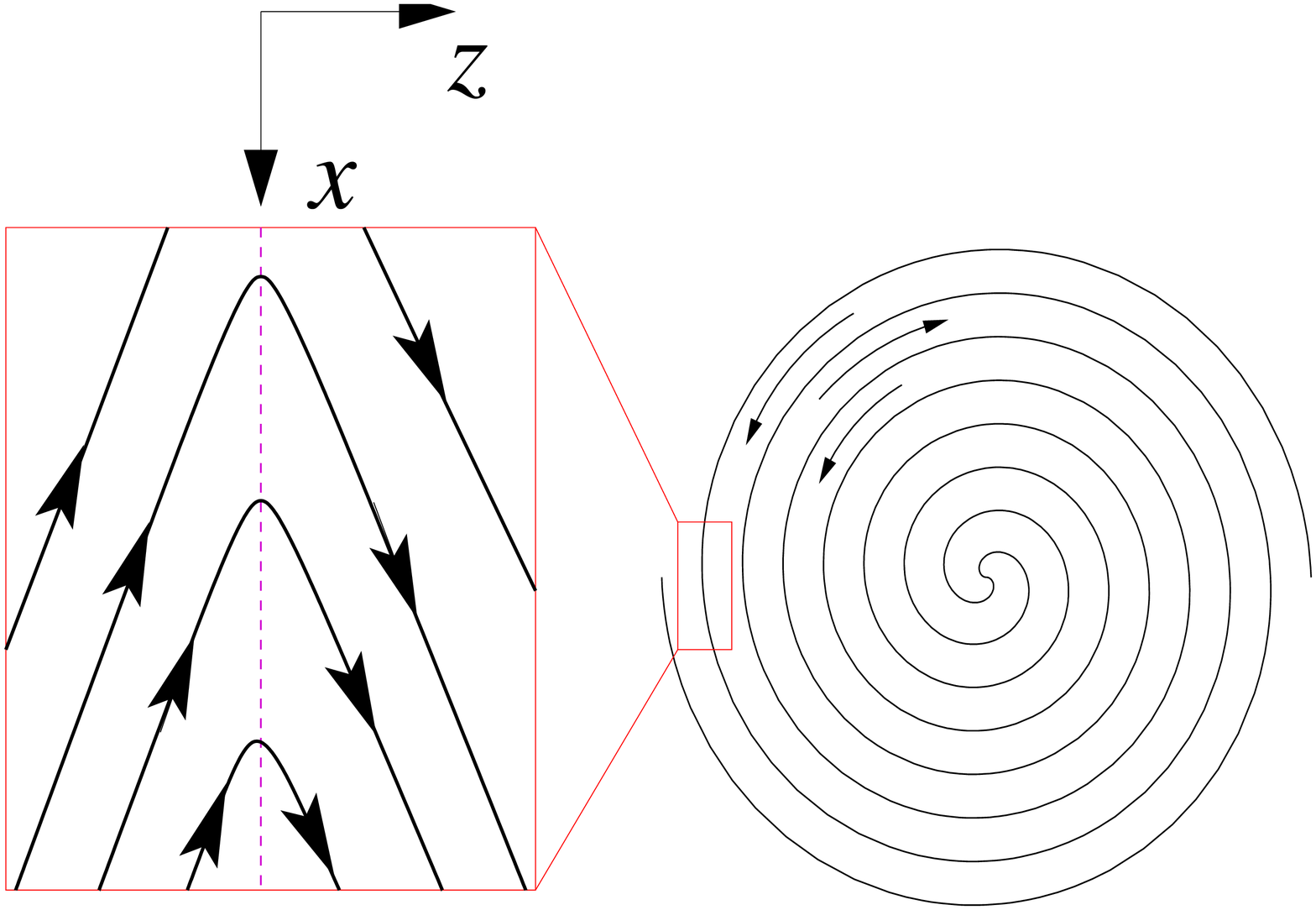}
\end{center}
\end{figure}
The value of $\sigma$ in the 
energetically less important, polar parts of the outflow is
uncertain. \citet{delzannaetal04} concentrate on the case in which $\sigma$ is
independent of angle, with a current sheet in the equatorial
plane, as in the case of the \lq\lq aligned\rq\rq\ split monopole. On the
other hand, \citet{komissarovlyubarsky03,komissarovlyubarsky04} are motivated
by the scenario where low magnetisation is achieved
by complete 
dissipation of the oscillating component of the magnetic field in the 
\lq\lq striped wind\rq\rq\ of the oblique split-monopole 
solution \citep{coroniti90,lyubarskykirk01}. 
On the equator, the dc component of the 
field in this solution 
vanishes for all obliquities, and so $\sigma=0$ here after dissipation
of the stripes. 
But, for all except the exactly perpendicular case, the stripes are confined
to a band of latitudes around the equator, implying a high
value of $\sigma$ close to the poles. Interestingly, both groups agree
that the average value of $\sigma$, i.e., the ratio of the total luminosity in
Poynting flux to the total in particle-born flux, should be a few percent.

\begin{table}
\caption{\label{accelerationtable}
Dependence of the bulk Lorentz factor $\Gamma$ 
of the pulsar wind on radius 
for different dissipation mechanisms. $r_{\rm max}$ is the radius at which
dissipation is complete, $r_{\rm L}$ is the radius of the light-cylinder,
$\mu=L/(\dot{N}_\pm mc^2)$ is Michel's parameter (with $L$ the 
spin-down luminosity, and $\dot{N}_\pm$ the pair loss rate in the wind) and
$\hat L=L\pi^3 e^2/(m^2c^5\Omega_{\rm w})$ 
is a dimensionless measure of the 
spin-down luminosity divided by the solid angle 
$\Omega_{\rm w}$ occupied by the wind
($\hat L\approx1.5\times10^{22}$ for the Crab pulsar). 
$\hat L$  is 
approximately equal to the square of
the pole to equator potential in units of the electron rest-mass.}
\begin{center}
\begin{tabular}{lll}
\hline
Slow dissipation   & Tearing-mode & Fast dissipation \\
\hline
$\Gamma\propto r^{1/2}$&$\Gamma\propto r^{5/12}$&$\Gamma\propto r^{1/3}$\\
${r_{\rm max}/ r_{\rm L}}\,=\,\hat{L}^{1/2}$
&${r_{\rm max}/r_{\rm L}}\,=\,\mu^{4/5}\hat{L}^{3/10}$
&${r_{\rm max}/ r_{\rm L}}\,=\,\mu^2$\\
\hline
\end{tabular}
\end{center}
\end{table}

\section{The $\sigma$ paradox}
The exact
solution with  $\sigma\rightarrow\infty$
found by \citet{michel73} has radial poloidal field. 
Its large $\sigma$ counterpart
also appears neither to collimate significantly, nor to accelerate
\citep{bogovalovtsinganos99}. Thus, if, as expected, 
$\sigma$ is large close to the star, it should continue to be so 
out to the termination shock. The small magnetisation 
implied by observations completes the $\sigma$ paradox.

There are several possible escape routes:
\begin{enumerate}
\item
\label{mitch}
Before the advent of the high-resolution images of the Crab Nebula, 
\citet{begelman98} suggested that $\sigma$ remains large even outside the
termination shock, and that the magnetic energy dissipates in the Nebula. 
However, this now seems to be ruled out by the good agreement of 
low $\sigma$, ideal MHD
simulations and observations in the optical and X-ray bands.
\item
\label{juri}
\citet{lyubarsky03b} also proposed that $\sigma$ remains large in the wind, but
that the energy carried in the oscillating component of the magnetic field
dissipates in a transition region that can be considered as a 
special  kind of termination shock. Such a configuration leads to the same
initial conditions for the MHD simulations as does a low $\sigma$ wind, 
provided the transition region remains thin. 
This is an interesting suggestion that
needs further investigation. In particular, it would be important to identify 
any spectral signature that could depend on the position of the termination
shock. 
\item
\label{nektarios}
Some collimation and, therefore, acceleration and conversion of magnetic into
particle-born energy might be possible if the initial distribution of poloidal
flux is sufficiently anisotropic \citep{chiuehlibegelman98,vlahakis04}. 
However, whether this is a realistic possibility can only be determined by 
constructing 
a model of the inner magnetosphere.  
\item
\label{damping}
\citet{coroniti90} and \citet{michel94} proposed that the oscillating
component of the magnetic field in the striped wind picture 
should gradually damp by dissipation of the
magnetic energy at the field line reversals. However, these early papers did
not include the associated acceleration of the wind and 
the resulting dilation of the
dissipation timescale. 
\end{enumerate}

Routes (\ref{juri}) and (\ref{nektarios}) 
await further theoretical development,
and I now turn to a discussion of route (\ref{damping}).

\section{Dissipation mechanisms}
To assess the importance of wave damping as a possible solution of the
$\sigma$ problem it is first necessary to identify the nature of the wave. 
Close to the pulsar, electromagnetic modes do not propagate 
\citep{usov75,melatosmelrose96}. On the other hand, 
both the fast magnetosonic wave \citep{lyubarsky03a} and the entropy wave
\citep{lyubarskykirk01} are possible. In principle, these may convert 
spontaneously into electromagnetic modes at larger radius \citep{melatos98},
but the conditions under which this happens have 
not yet been fully investigated. If the waves persist
as subluminal modes, damping is inevitable \citep{usov75}.
 
The rate of damping of the striped wind pattern and, therefore, the 
viability of escape route (\ref{damping}), 
depends on 
the speed with
which magnetic reconnection can proceed in a relativistic current 
sheet. These
structures (see Fig.~\ref{prlfigure}) 
differ from their more familiar counterparts in solar and
magnetospheric physics because they do not permit evacuation of the heated
plasma from the sheet as in a Sweet-Parker or Petschek-type 
reconnection model 
\citep{kirk04}. 
Instead, 
either the sheet steadily thickens, or plasma is
ejected along the induced electric field (the $\vec{y}$-direction in
Fig.~\ref{prlfigure}). 
As dissipation proceeds, the current sheet separating regions of opposite
magnetic polarity fills with hot plasma. If this process proceeds
slowly (not faster than the timescale needed for a fluid element in 
the wind to double its radius) a small-wavelength approximation can be used to
analyse the evolution of the stripes. On the other hand, rapid dissipation 
cannot be ruled out, and is difficult to quantify. 

Concentrating on the small-wavelength approximation, the physical picture is
one in which 
the sheet remains in approximate pressure
equilibrium with the surrounding magnetic field, which is dragged into it and
annihilated. Spherical symmetry must be assumed to make the analysis 
tractable, but since the flow is highly
relativistic, pressure imbalance in
the $\theta$ direction is unlikely to be important, and we can assume the
solution holds independently along each radius vector. 
A lower limit on the rate at which annihilation can proceed is
found by arguing that the sheet width cannot be smaller than the gyro radius
of a \lq\lq hot\rq\rq\ particle 
in the confining field \citep{lyubarskykirk01}. However, this 
rate is too slow to be of interest in the case of the Crab. A
more realistic estimate of the annihilation rate follows 
by assuming it equals the 
timescale on which the relativistic tearing-mode instability operates
\citep{lyubarsky96,kirkskjaeraasen03}. But, again, this is not fast enough
to resolve the $\sigma$ paradox for the Crab. An upper limit, subject to the
assumptions inherent in this method, is given by demanding that the hot sheet
expand at most sonically. This {\em does} 
provide a viable escape route for the case of the Crab, 
but only if the number of electron/positron pairs carried
from the pulsar 
by the wind is significantly larger ($>3\times10^{40}\,\textrm{s}^{-1}$)
than the rate found in standard 
pair-production calculations \citep{hibschmanarons01a,hibschmanarons01b}. 
Independent support for 
injection of pairs into the Nebula at this high rate 
is provided by models of the radio to hard X-ray synchrotron spectrum 
\citep{gallantetal02}, so that we may still have much to learn about
where and how a pulsar produces electron/positron pairs.

For each of these proposed dissipation processes, 
the acceleration of the wind can be described by a similarity solution 
in which the bulk Lorentz factor is proportional to a power of the radius. 
These results are summarised in Table~\ref{accelerationtable}. 
In principle, they make possible a prediction of both the average $\sigma$
and the Lorentz factor of a pulsar wind as a function of radius. 
Given the inclination angle of the magnetic dipole axis, predictions can also
be made of the dependence of these quantities on colatitude.  

\section{Observing the wind}
\label{observability}

In the case of gamma-rays, an observational upper limit on the wind
emission yields a firm constraint on the pulsar wind.  
Even if the plasma remains completely cold, if its bulk Lorentz
factor is large, it will radiate
by Compton upscattering the ambient target
photons. 
Using thermal X-rays from the surface of the Crab pulsar as targets, 
\citet{bogovalovaharonian00} interpreted 
the detected unpulsed flux in TeV gamma-rays as an upper limit 
on the contribution of the wind,  
and derived a {\em lower} limit 
of $5r_{\rm L}$ on the radius at
which the Lorentz factor of the Crab wind can reach $10^6$. This constraint
is satisfied by all the models presented in Table~\ref{accelerationtable}.

In the optical and X-ray images the region around the Crab pulsar appears dark 
out to a distance of about $12\,\textrm{arcsec}$. 
However, this does not mean the wind
does not radiate, but merely that along a line of sight that passes between
about $0.5\,\textrm{arcsec}$ and $12\,\textrm{arcsec}$ of the pulsar, the
emission of the wind is beamed away from the observer. 
This occurs quite naturally in a relativistic, radial wind.
If the bulk Lorentz factor is proportional to $r^q$ with $q<1$ ---
as in the cases considered in Table~\ref{accelerationtable} --- 
the most stringent constraint is imposed at the termination shock.
For the emission to be beamed away from the observer at that radius
requires $\Gamma>12\,\textrm{arcsec}/0.5\,\textrm{arcsec}=24$. 
This is much smaller
than the Lorentz factor usually assumed for the plasma entering  
the termination shock. 
Consequently, even if it radiates at all radii, the wind 
should appear to contribute only to the point source, leaving the surrounding
region dark.

An analogous argument can be advanced concerning pulsation of the wind
emission. If the striped pattern gives rise to a phase dependence of the
emissivity, for example, to increased emission at the field reversals, the
radiation from radius $r$ will appear pulsed at the rotation period of 
the star provided $r/r_{\rm L}<\Gamma^2$ \citep{kirkskjaeraasengallant02}. 

Has this radiation been observed? 
To answer this question, one must be able to
untangle the contribution of the stellar surface, magnetosphere and wind, all
of which are co-spatial (and point-like) when observed with current 
instruments. This is a difficult task, especially
given the uncertainties associated with modelling the pulsed
emission. However, one characteristic of synchrotron emission from the wind
that does not appear to be shared by any magnetospheric model 
\citep{dykshardingrudak04,kaspirobertsharding04}
is the direction
of linear polarisation of the off-pulse or \lq\lq dc\rq\rq\ component. 
In the striped wind, the
polarisation vector of radiation emitted close to the field 
reversals is not well constrained,
but outside of these potentially pulse-producing regions, 
the magnetic field is purely toroidal. This predicts that the
electric vector of the dc component of synchrotron radiation should lie along
the projection onto the sky of the rotation axis of the neutron star. In the
case of the Crab pulsar, the optical emission appears to have precisely this
property \citep{kellner02}.    

\ack{
I thank Gottfried Kanbach for discussions of the optical emission of
the Crab pulsar
}


\end{document}